\documentclass[aps,prd,amsmath,amssymb,superscriptaddress,eqsecnum,reprint,showpacs]{revtex4-1}
\usepackage{color}
\usepackage[pdftex]{graphicx}
\usepackage[alwaysadjust]{paralist}
\setdefaultenum{(1)}{(a)}{(i)}{(A)}
\newcommand{\dropme}[1]{\textcolor{red}{{}}}

\newcommand{\leaveout}[1]{}
\newcommand{\fixme}[1]{}

\newcommand{\Dg}{\Delta g_{\ell}}
\newcommand{\g}{g_{\ell}}
\newcommand{\gt}{g_{\ell}}

\newcommand{\DGw}[3]{\Delta G_{\ell}(#1,#2;#3)}

\newcommand{\f}{f_{\ell}}

\newcommand{\an}[1]{a_{#1}}

\newcommand{\nb}{\nu}

\newcommand{\W}[1]{W\left(#1\right)}
\newcommand{\wbQNM}{\ob_{\ell n}}
\newcommand{\ob}{\omega}
\newcommand{\rb}{r}

\newcommand{\psio}{\tilde \psi_1}
\newcommand{\psit}{\tilde \psi_2}

\newcommand{\A}{i\alpha_1^2}

\newcommand{\En}{A_1}
\newcommand{\Fn}{B_1}
\newcommand{\Gn}{A_2}
\newcommand{\Hn}{B_2}
\newcommand{\Ai}{A_i}
\newcommand{\Bi}{B_i}

\newcommand{\ept}{e^{\rnu}}
\newcommand{\emt}{e^{-\rnu}}
\newcommand{\cf}{c_f(\nu)}

\newcommand{\iQ}{k} 

\newcommand{\az}{\alpha_0}
\renewcommand{\aa}{\alpha_1}
\newcommand{\ab}{\alpha_2}
\newcommand{\ac}{\alpha_3}

\newcommand{\ind}{k}

\newcommand{\rnu}{v}

\newcommand{\suma}{{\sum}}

\begin{document}


\author{Marc Casals}
\email{mcasals@perimeterinstitute.ca; 
marc.casals@ucd.ie}
\affiliation{Perimeter Institute for Theoretical Physics, Waterloo, Ontario, Canada N2L 2Y5}
\affiliation{Department of Physics,
University of Guelph,
Guelph, Ontario, Canada N1G 2W1}
\affiliation{School of Mathematical Sciences and Complex \& Adaptive Systems
Laboratory, University College Dublin, Belfield, Dublin 4, Ireland}

\author{Adrian Ottewill}
\email{adrian.ottewill@ucd.ie}
\affiliation{School of Mathematical Sciences and Complex \& Adaptive Systems
Laboratory, University College Dublin, Belfield, Dublin 4, Ireland}

\title{Spectroscopy of the Schwarzschild Black Hole at Arbitrary Frequencies}

\begin{abstract}

Linear field perturbations of a black hole are described by  
the Green function of the wave equation that they obey.
After Fourier decomposing the Green function, its
two natural contributions
are
given by poles (quasinormal modes) and a largely unexplored branch cut in the complex-frequency plane.
We present new analytic methods for calculating  
 the branch cut on a Schwarzschild black hole for {\it arbitrary} values of the frequency.
The branch cut yields a power-law tail decay for late times in the response of a black hole to an initial perturbation.
We determine explicitly the first  three orders  in the power-law and show that the branch cut also yields a new logarithmic behaviour
$T^{-2\ell-5}\ln  T$ for late times.
Before the tail sets in, the quasinormal modes dominate the black hole response.
For electromagnetic perturbations, the quasinormal mode frequencies approach the branch cut at large overtone index $n$. 
We 
determine
these 
frequencies up to 
$n^{-5/2}$ and,  formally, to {\it arbitrary} order.
Highly-damped quasinormal modes are of particular interest in that they have been linked to quantum properties of black holes.
\end{abstract}

\maketitle




The retarded Green function for linear field perturbations in black hole spacetimes is 
of central physical importance in classical and quantum gravity.
%
An understanding of the make-up of the Green function is obtained by performing a  Fourier transform, thus yielding an integration
just above the real-frequency ($\omega$) axis.
In his seminal paper, Leaver~\cite{Leaver:1986} deformed this  real-$\omega$ integration 
in the case of Schwarzschild spacetime 
 into a contour on the complex-$\omega$ plane.
He thus unraveled three contributions making up the Green function: (1) a high-frequency arc, (2) a series over poles 
of the Green function (quasinormal modes QNMs), and (3) 
an integral of modes around a branch cut originating at $\omega=0$ and extending down the negative imaginary axis (NIA),
which we refer to as branch cut modes (BCMs).
The three contributions dominate the black hole response to an initial perturbation at different time regimes.
The high-frequency arc 
yields a `direct' contribution which 
is expected to vanish
after
a certain finite time~\cite{Ching:1994bd,Ching:1995}.

The QNM contribution to the  Green function dominates
the black hole response
during `intermediate' times and it has been extensively investigated (e.g.,~\cite{Berti:2009kk} for a review). 
At `late times' the QNM contribution decays exponentially, with a decay rate given by the overtone number
$n\in\mathbb{Z}^+$.
QNMs have also triggered numerous interpretations in different contexts in classical and quantum physics, 
ranging from astrophysical `ringdown'~\cite{Vishveshwara:1970zz} 
to
Hawking radiation~\cite{York:1983,Keshet:2007be}, 
the `gauge-gravity duality' (\cite{Horowitz:1999jd} for Schwarzschild black holes which are asymptotically anti-de Sitter and~\cite{Bertini:2011ga} for 
asymptotically flat ones),
black hole area quantization~\cite{Bekenstein:1974jk,Bekenstein:1995ju,Hod:1998vk,Dreyer2003,Maggiore:2007nq} and
structure of spacetime at the shortest length scales~\cite{Babb:2011ga}. 
The quantum interpretations are given in the highly-damped limit, i.e, for large $n$.
The highly-damped QNM frequencies in Schwarzschild 
have been calculated up to next-to-leading order in~\cite{Motl&Neitzke,Neitzke:2003mz,MaassenvandenBrink:2003as,Musiri:2003bv,Musiri:2007zz}.
Despite all the efforts, the leading order of the real part of the frequencies for
electromagnetic perturbations
has remained elusive (only in~\cite{Cardoso:2003vt} they find numerical indications
that it goes like $n^{-3/2}$). 

The contribution from the BCMs, on the other hand, remains largely unexplored.
The technical difficulties of  its analysis mean
that most of the studies have been constrained to large radial coordinate as well as
small $\nu\equiv i\omega>0$ along the NIA. 
An exception is a large-$\nu$ asymptotic analysis of the BCMs in~\cite{MaassenvandenBrink:2003as} 
(and 
near  the algebraically-special frequency in~\cite{MaassenvandenBrink:2000ru}) solely for
gravitational perturbations.
The small-$\nu$ BCMs are known to give rise to a power-law tail
 decay at `late' times of an initial perturbation~\cite{Price:1971fb,Price:1972pw,Leaver:1986,Ching:1994bd}.
In general, however, there is an appreciable time interval between when the QNM contribution becomes
 negligible and when the power-law tail starts~\cite{Andersson:1997}.
The calculation of the BCMs for general values of the frequency (i.e., not in the asymptotically
small nor large regimes), to the best of our knowledge has only been attempted in~\cite{Leung:2003ix,Leung:2003eq}
where the radial functions \fixme{maybe not $\f$?}
were calculated 
off
the NIA via a numerical integration of the radial ODE (\ref{eq:radial ODE})
followed by extrapolation to the NIA\fixme{this is not clear?}, and only for the gravitational case.

In this Letter we present 
the following new results:
\begin{compactenum}
\item 
A new analytic method for the 
calculation of the BCMs directly {\it on} the 
NIA
 and valid for {\it any} value of $\nu$.
In particular, this method provides analytic access for the first time to the `mid'-$\nu$ regime.
\item 
%
A consistent expansion up to $4$th order for small-$\nu$ of the BCMs for arbitrary value of the radial coordinate. 
We explicitly derive a new logarithmic behaviour
$T^{-2\ell-5}\ln  T$ 
at late times.
\item 
%
A large-$\nu$ asymptotic analysis of the BCMs. 
It
shows a formal divergence, which is expected to be cancelled out by the other contributions to the Green function.
\item 
%
A new asymptotic analysis for large-$n$ of the electromagnetic QNMs. The analysis is formally valid up to {\it arbitrary}
order in $n$; 
we explicitly calculate the corresponding frequencies
 up to $n^{-5/2}$.
\end{compactenum}
Methods in (1)--(3) provide the first full analytic account of the BCMs and 
they are valid for
any  
spin $s=0$ (scalar), $1$ (electromagnetic) and $2$ (gravitational) of the field perturbation.
For the QNM calculation we focus on  spin-1 as this is the least well understood case.
We note that 
spin-1
 perturbations are acquiring increasing 
importance~\cite{Schnittman:2010wy,Tamburini:2011tk}, 
although it is expected that only the lowest overtones of the QNMs are astrophysically relevant.

\dropme{
Details of these calculations will be presented in extended papers~\cite{Casals:Ottewill:2011largeBC} and~\cite{Casals:Ottewill:2012SmallMidBC}. }
We present details in~\cite{Casals:Ottewill:2012SmallMidBC} and~\cite{Casals:Ottewill:2011largeBC}.
We take units $c=G=2M=1$, where $M$ is the mass of the black hole.

\section{The Green function \& branch cut}
After carrying out a Fourier transform and a multipole decomposition, 
the radial and time parts of the retarded Green function for
linear fields on a Schwarzschild black hole can be written as
\begin{align} \label{eq:Green}
&
G^{ret}_{\ell}(r,r';t)\equiv \!\!\!\!\!
\int\limits_{-\infty+ic}^{\infty+ic} \!\!\!\frac{d\omega}{2\pi}\, \frac{\f(r_<,\omega)\g(r_>,\omega)}{\W{\omega}}e^{-i\omega t}
\end{align}
where $c>0$, $\ell$ is the multipole number, $r_>\equiv \max(r,r')$, $r_<\equiv \min(r,r')$
and $\W{\omega}$ is the Wronskian of the two functions $\f$ and $\g$.
These functions are linearly independent solutions of the radial ODE
\begin{align}
\label{eq:radial ODE}
&\left\{\frac{d^2}{dr_*^2}+\omega^2-\left(1-\frac{1}{r}\right)\left[\frac{\lambda }{r^2}+\frac{(1-s^2)}{r^3}\right]\right\}
\psi_{\ell}
=0
\end{align}
where $r_*\equiv r+\ln(r-1)$ and $\lambda\equiv \ell (\ell+1)$.
The solutions are uniquely determined when $\text{Im}(\ob)\ge 0$ by the
boundary conditions: $\f\sim e^{-i\omega r_*}$ as $r_*\to -\infty$ and $\g \sim  e^{+i\omega r_*}$ as $r_*\to \infty$.
The behaviour of the radial potential at infinity
 leads to a branch cut in the radial solution $\g$~\cite{Leaver:1986a,Ching:1995tj}.

The contour of integration in Eq.(\ref{eq:Green}) can be deformed in the complex-$\omega$ plane~\cite{Leaver:1986}
yielding a contribution from a high-frequency arc, 
a series over the residues (the QNMs) 
and a contribution 
from the branch cut along the NIA:
\begin{equation} \label{eq: G^BC integral}
G_{\ell}^{BC}(r,r';t)\equiv
\frac{1}{2\pi i}
\int\limits_{0}^{\infty } d\nu\ 
\DGw{r}{r'}{\nu}
e^{-\nu t},
\end{equation}
where the BCMs are
\begin{align} 
\label{eq:DeltaG in terms of Deltag}
&
\Delta G_{\ell}(r,r';\nu)\equiv
-\frac{2i\nu q(\nu)}{\left|W(-i\nu)\right|^2}\f(r,-i\nu)\f(r',-i\nu),
\end{align}
with $q(\nu)\equiv -i\Dg(r,\nu)/\g(r,i\nu)$
where  $\Dg(r,\nu)\equiv \lim_{\epsilon\to 0^+}\left[\g (r,\epsilon-i\nu)-\g (r,-\epsilon-i\nu)\right]$ is the discontinuity of $\g$ across the branch cut.




We present here methods for the analytic calculation of the BCMs.
We calculate $\f$  using the  Jaff\'e series, Eq.39~\cite{Leaver:1986a}.
The coefficients of this series, which we denote by  $\an{\ind}$, satisfy a 3-term recurrence relation.
We calculate  $\gt$ using the  series in Eq.73~\cite{Leaver:1986a}, which is in terms of 
 the confluent hypergeometric $U$-function and the coefficients $\an{\ind}$.
This series has  seldom been used and one must be aware that, in order for $\g$ to satisfy the correct boundary condition,
we must
set
 $\an{\ind=0}=(-2i\ob)^{s+1-2i\ob}$, which itself has a branch cut.
To find an expression for $\Delta G_{\ell}$ {\it on} the NIA we exploit this series
by combining it with the known behavior of the $U$-function across its branch cut:
\begin{align} 
\label{eq:Leaver-Liu series for Deltagt}
&
\Dg(r,\nu)=\frac{\rb^{1+s}e^{-\nu r}}{\left(\rb-1\right)^{\nb}}
\frac{2\pi ie^{\pi i (s+1-2\nb)}}{\Gamma(1-2\nb)}\times  \\
&\quad \sum_{\ind=0}^{\infty}
\an{\ind}
\frac{(-1)^{\ind}\Gamma(1+\ind-2\nb)U(s-\ind+2\nb,2s+1,2\nu r)}{\Gamma(1+s+\ind-2\nb)\Gamma(1-s+\ind-2\nb)}  \nonumber
\end{align}
where we are taking the principal branch both for $\an{\ind=0}$ and for the $U$-function.
In order to check the convergence of this series, we require the behaviour for large-$\ind$ of the coefficients $\an{\ind}$.
Using the  {\it Birkhoff series} as in App.B~\cite{Wimp}, we find the leading order
$\an{\ind} \sim \ind^{-\nb-3/4}e^{\pm 2\sqrt{2\nb \ind}i}$ (we have calculated up to four orders higher
in~\cite{Casals:Ottewill:2012SmallMidBC}) as $k\to \infty$.
We note that this behaviour corrects Leaver's Eq.46~\cite{Leaver:1986a} in the power `$-\nb$' instead of `$-2\nb$'.
The integral test then shows that the series (\ref{eq:Leaver-Liu series for Deltagt}) converges  for any $\nb>0$.
Although  convergent, the usefulness of (\ref{eq:Leaver-Liu series for Deltagt}) at small-$\nb$
 is limited since convergence becomes slower
as $\nb$ approaches 0 while,
 for large-$\nb$, 
$\Delta G_{\ell}$
grows and oscillates for fixed $r$ and $r'$.
Therefore we complement our analytic method with asymptotic results for small and large $\nu$.

The small-$|\ob|$ asymptotics are based on 
an extension of the MST formalism~\cite{Mano:Suzuki:Takasugi:1996,Sasaki:2003xr}.
We start with the ansatz
\begin{align}  \label{eq:f small-nu}
   &
\f =
\frac{e^{-i\ob\rb}(\rb-1)^{-i\ob}}{ \sum_{j=-\infty}^{\infty}a_j^{\mu}}\times\\
  & \sum_{k=-\infty}^{\infty}a_k^{\mu} \frac{\Gamma(k+\mu+s+1-i\ob)\Gamma(-k-\mu+s-i\ob)}
{\Gamma(1-2i\ob)}\times\nonumber\\
&{}_2F_1(k+\mu+s+1-i\ob,-k-\mu+s-i\ob;1-2i\ob;1-\rb).
\nonumber
\end{align}
 Imposing Eq.(\ref{eq:radial ODE}) yields a 3-term recurrence relation for $a_k^{\mu}$ and requiring convergence as $k\to\pm\infty$ yields an equation for $\mu$, 
 that may readily be solved perturbatively in $\omega$ from starting values
$\mu_{\omega=0}=\ell$ and $\mu_{\omega=0}=-\ell-1$.
Likewise for the coefficients $a_k^{\mu}$, taking $a_0^{\mu}=1$ we obtain
\begin{align} 
\label{eq:a_k^mu}
\nonumber
&
a_{1}^{\mu}= \frac{(\ell+1-s)^2}{2(\ell+1)(2\ell+1)} \left[-i \omega + \frac{\omega^2}{\ell+1}+O\left(\omega^3\right)\right]
\\ &
 a_{2}^{\mu}=
-\frac{(\ell+1-s)^2(\ell+2-s)^2}{4(\ell+1)(2\ell+1)(2\ell+3)^2}\omega^2+O\left(\omega^3\right)
\end{align}
while $a_{-1}^\mu$ and $a_{-2}^\mu$ are given by the corresponding terms with $\ell\to-\ell-1$.
(Apparent possible singularities in these coefficients are removable.)
%
%

The $k=0$ term in Eq.(\ref{eq:f small-nu}) corresponds to Page's 
Eq.A.9~\cite{ar:PageI'76}.
To obtain higher order aymptotics we employ the Barnes  integral representation
of the hypergeometric functions \cite{bk:Erdelyi1} which involves a contour in the complex $z$-plane from $-i \infty$ to $i \infty$
threading between the poles of $\Gamma(k+\mu+s+1-i\ob+z)$, $\Gamma(-k-\mu+s-i\ob+z)$ and $\Gamma(-z)$.
As $\omega\to0$ double poles arise at the non-negative integers from 0 to $\max(k+\ell-s,-k-\ell-1-s)$, however
we may move the contour to the right 
of all these ambient double poles picking up polynomials in $r$ with 
coefficients readily expanded in powers of $\omega$, leaving a regular  contour which admits immediate expansion in powers of~$\nu$.

By the method of MST we can also construct $\g$ and hence determine 
$q(\nu)$ and $W$. 
For compactness, we only give the following small-$\nu$ expressions 
for the case $s=0$ (cases $s=1$ and $2$ are presented in~\cite{Casals:Ottewill:2012SmallMidBC}), 
\leaveout{
\begin{widetext}
\begin{align}
&
q_3\equiv 
\frac{- \pi  2^{1-2 \ell} \Gamma^2(\ell+1)}{\Gamma^4 (2 \ell+2)}
    \left\{-4 (2 \ell+1)^2 \left(-8 \left(H_{\ell}\right){}^2+8 H_{\ell}+3 H_\ell^{(2)}+
   2H_{\infty}^{(2)}
    \right)+
\right. \nonumber  \\ &\left.
   (2 \ell+1) \left(\frac{1}{(2 \ell-1) (2 \ell+3)}+15\right) \left(H_{\ell}-4 H_{2 \ell}+\log (4 \nu )+\gamma_E
   \right)+\frac{4 \ell}{(2 \ell-1) (2 \ell+3)}+\frac{4 (2 \ell+1)}{(2 \ell-1)^2 (2 \ell+3)^2}+32 \ell^2+62 \ell-7\right\}
\end{align}
\end{widetext}
}
\begin{widetext}
\begin{align}\label{q/W^2 s=0 gral l}
&
\frac{q(\nu)}{|W|^2}= -\frac{ (-1)^{\ell}\pi}{2^{2\ell-3}}\left(\frac{(2\ell+1) \ell!}{\left( \left(2\ell+1\right)!!\right)^2}\right)^2 \left[\nu^{2\ell+1} -
\nu^{2\ell+2}\left(\frac{-32 \ell^3-63 \ell^2-7 \ell+23}{2(2\ell+3)(2\ell+1)(2\ell-1)}+4 H_{\ell}\ \right)\right] \nonumber\\
&
-\frac{ (-1)^{\ell} \pi} { 2^{2 \ell-1}} \left(\frac{(2\ell+1)\ell!}{((2 \ell+1)!!)^2}\right)^2 \nu^{2\ell+3}
    \left[\frac{4(15\ell^2+15\ell-11)}{(2 \ell-1)(2\ell+1) (2 \ell+3)} \left(\ln (2 \nu )+H_{\ell}-4 H_{2 \ell}+
\gamma_E
   \right)\right.   \\ &\left.
-4  \left(-8 H_{\ell}{}^2+8 H_{\ell}+3 H_\ell^{(2)}+
   2H_{\infty}^{(2)}
    \right)+
  \frac{512\ell^6+2016 \ell^5 +1616 \ell^4-1472 \ell^3- 1128 \ell^2+722 \ell-59}{(2 \ell-1)^2  (2 \ell+1)^2 (2 \ell+3)^2}\right] +o(\nu^{2\ell+3})
   \nonumber
\end{align}
\end{widetext}
where $H^{(r)}_{\ell}$ is the $\ell$-th harmonic number of order $r$.
We note that the $\ln\nb$ term  at second-to-leading order originates both in $q(\nu)$ and in
$W$.
In fact, both functions possess a $\ln\nb$ already at next-to-leading order for small-$\nb$, but they cancel
each other out in $q/|W|^2$.
Similarly, the coefficient of a potential term in $q/|W|^2$ of order $\nb^{2\ell+3}(\ln\nb)^2$ is actually zero.

\begin{figure}[t]
\begin{center}
                \includegraphics[width=\columnwidth]{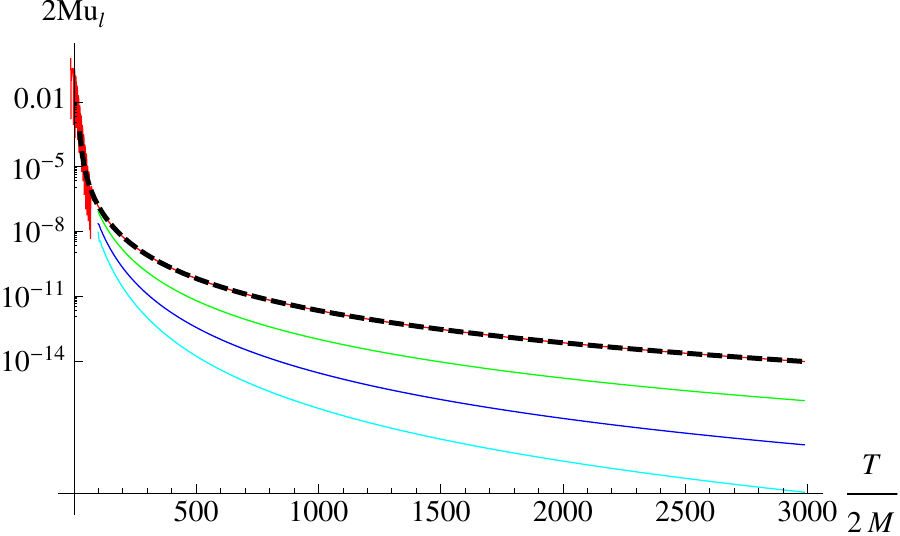}
\end{center}
\vspace{-4mm}
\caption{(Color online). Full perturbation response 
$u_{\ell}$ to the Gaussian described above Eq.(\ref{eq:pert_asymp}) compared to the late-time asymptotics.
Solid-red: numerical solution; dashed-black: Eq.(\ref{eq:pert_asymp}); lower curves: numerical solution minus the first (green), first 2 (blue) and first 4 (cyan) terms in 
Eq.(\ref{eq:pert_asymp}).
}
\label{fig:perturbation response}
\end{figure}

Let us now investigate the branch cut contribution to the black hole response to an initial perturbation given by the
field $u_{\ell}^{ic}$ and its time derivative $\dot u_{\ell}^{ic}$ at $t=0$:
\begin{align}\label{eq:perturbation BC}
&
u_{\ell}^{BC}(r_*,t)\equiv 
\\ &
\int_{-\infty}^{\infty}dr'_*\left[  G_{\ell}^{BC}(r,r';t)\dot u_{\ell}^{ic}(r'_*)+u_{\ell}^{ic}(r'_*)\partial_t  G_{\ell}^{BC}(r,r';t)\right]
\nonumber
\end{align}
We obtain the asymptotics of the response for late times $T\equiv t-r_*-r'_*$ 
using 
Eqs.(\ref{eq:DeltaG in terms of Deltag}) and (\ref{eq:f small-nu})--(\ref{q/W^2 s=0 gral l}).
We note the following features. 
The orders $\nb^{2\ell+2}$ and $\nb^{2\ell+3}$ in the BCMs $\Delta G_{\ell}$ yield
tail terms behaving like $T^{-2\ell-3}$ and  $T^{-2\ell-4}$, respectively.
We have thus generalized Leaver's Eq.56~\cite{Leaver:1986} to finite values of $r$.
Furthermore, Eq.56~\cite{Leaver:1986} is an expression containing the leading orders from $u_{\ell}^{ic}$
and from $\dot u_{\ell}^{ic}$. However, the next-to-leading order from $\dot u_{\ell}^{ic}$ will be of the same order
as the leading-order from $u_{\ell}^{ic}$. In our approach above we consistently give a series in small-$\nb$,
thus obtaining the correct next-to-leading order term for large-$T$ in the power-law tail.
Importantly, 
we also obtain the following two
orders 
in the perturbation response: $T^{-2\ell-5}\ln T$ and $T^{-2\ell-5}$.
We note the interesting $T^{-2\ell-5}\ln T$ behaviour,
 which is due to the $\nb^{2\ell+3}\ln\nb$ term
 in Eq.(\ref{q/W^2 s=0 gral l}).
To the best of our knowledge, this is the first time in the literature that any of the above features has been obtained.
The logarithmic behaviour is not completely surprising given the calculations in~\cite{Ching:1995tj}.
However, one may be led to a wrong logarithmic behaviour~\cite{Hod:2009my} if the calculations are not performed in detail.
In order to exemplify our results, we give the explicit asymptotic behaviour in the case $s=0$ and $\ell=1$
and initial data 
$u_{\ell}^{ic}=e^{-2\left(r_*-x_0\right)^2}$
and
$\dot u_{\ell}^{ic}=-4\left(r_*-x_0\right)u_{\ell}^{ic}$, with $x_0\equiv r_*(r=5)$.
The perturbation response due to the branch cut at $r=5$ at late times is given by
\begin{align}
\label{eq:pert_asymp}
u_{\ell}^{BC}&=
-2288.6\, T^{-5}+104770.5\, T^{-6}+
\\ &
\left[86968.2\ln T-3493893.9\right]T^{-7}
+o\left(T^{-7}\right).
\nonumber
\end{align}
Fig.~\ref{fig:perturbation response} shows that these  asymptotics are in excellent agreement with a numerical solution
of the wave equation.

At large-$\nb$,
we obtain the asymptotics
\begin{align}
\label{eq:large_nu}
&-\frac{2i\nu q(\nu)}{\left|W(-i\nu)\right|^2}
\sim
\begin{cases}
\dfrac{(-1)^{s/2} 2i \cos(2\pi\nb)}{\nu\left[1+3\cos^2(2\pi\nb)\right]},&\!\!\! s=0,2\\
\dfrac{-\sqrt{\pi}i\lambda\sin(2\pi\nb)}{\nb^{3/2}},&\!\!\!s=1
\end{cases} 
\\
&\f(r,-i\nu) \sim
\begin{cases}
(-1)^{s/2
}e^{\nu r_*}+\sin(2\pi\nb)e^{-\nu r_*},& s=0,2\\
\dfrac{\sqrt{\pi}\lambda}{2\nb^{1/2}\sin(2\pi\nb)}e^{\nu r_*}+e^{-\nu r_*},&s=1
\end{cases} 
\nonumber
\end{align}

These asymptotics show a divergence in $G_{\ell}^{BC}$ when $t<|r_*|+|r'_*|$.
They also lead to a divergence in the perturbation response at fixed $t$ and $r$ for a non-compact Gaussian as initial data.
Both types of divergences are expected to cancel out with the other contributions to the Green function.
We have thus provided a complete account of the BCMs for all frequencies along the NIA;
the behaviour is illustrated in Fig.\ref{fig:DeltaG s=l=2 and s=0,l=1}.


\begin{figure}[b!]
\begin{center}
                \includegraphics[width=8.2cm]{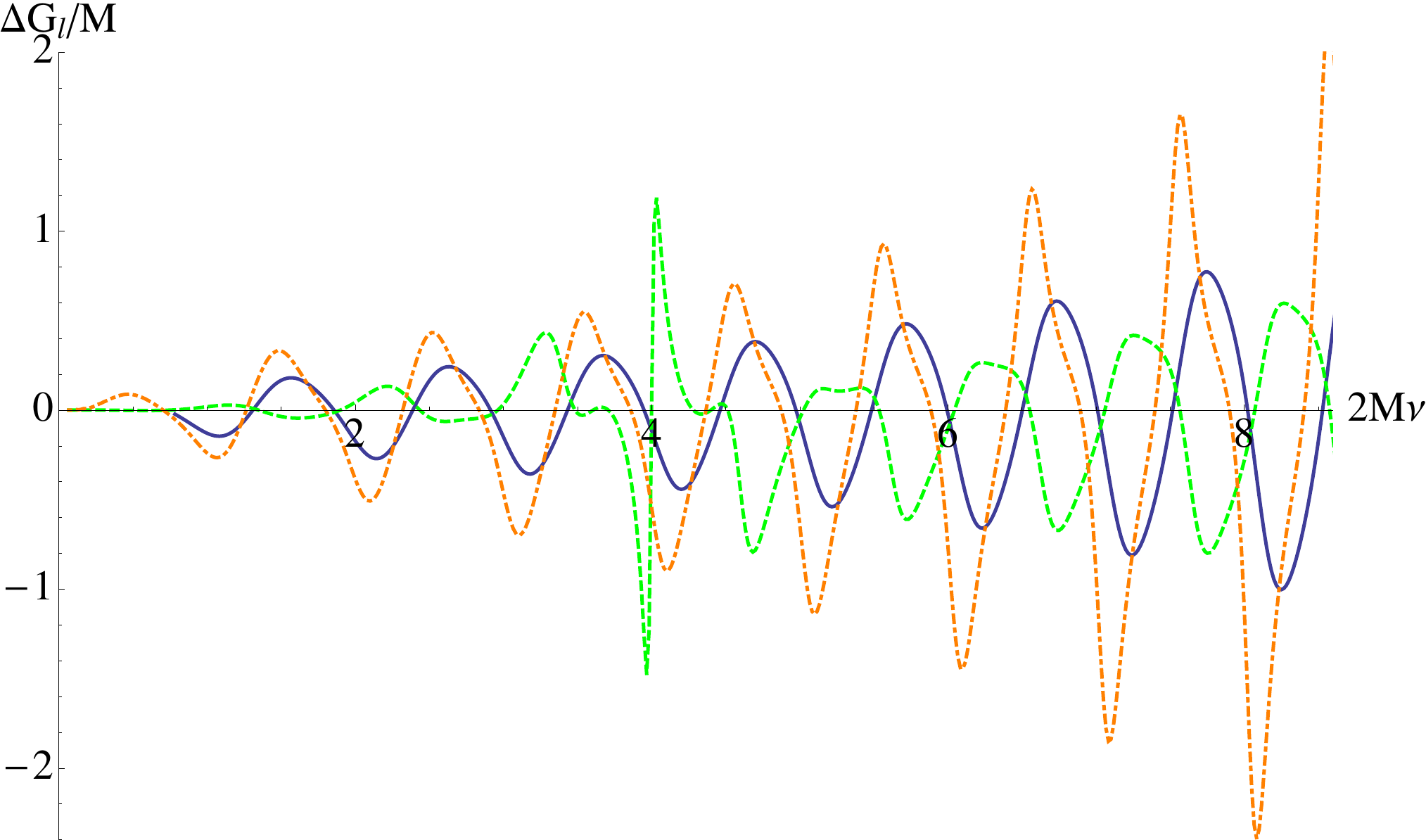}
\begin{tabular}{cc}
  \includegraphics[width=4cm]{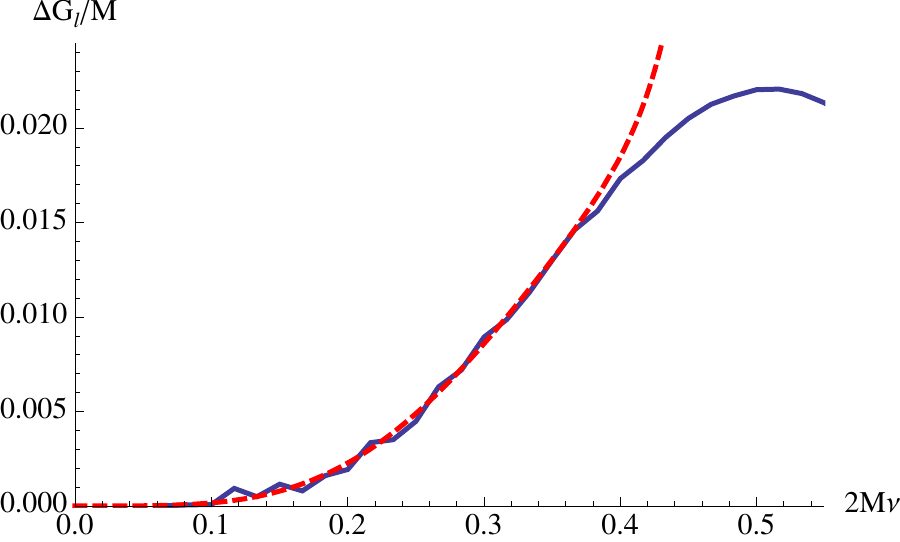}&
\includegraphics[width=4cm]{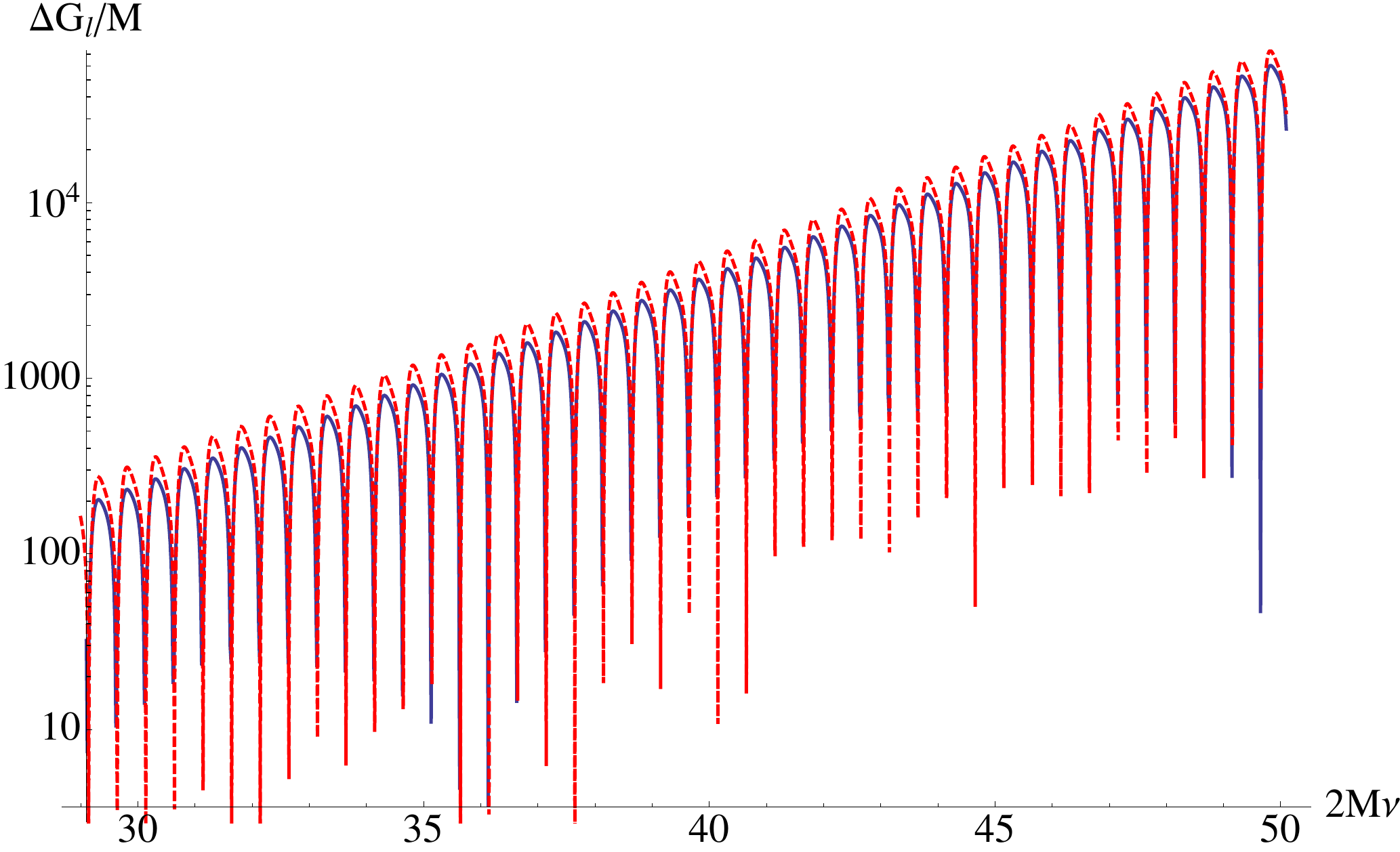}
\end{tabular}
\end{center}
\vspace{-4mm}
\caption{
(Color online).
$\Delta G_{\ell}$  as a function of $\nb$ for $r_*=0.1$ and $r'_*=0.2$.
(a) Using Eq.(\ref{eq:Leaver-Liu series for Deltagt}); dashed-green: $s=\ell=2$,  continuous-blue:  $s=0$, $\ell=1$, 
 dot-dashed-orange: $s=\ell=1$.
Note the interesting behaviour near the algebraically-special frequency~\cite{MaassenvandenBrink:2000ru} at $\nu=4$ for $s=2$.
(b) $s=0$, $\ell=1$ for 
small $\nb$;
continuous-blue using Eq.(\ref{eq:Leaver-Liu series for Deltagt}); dashed-red using 
Eq.(\ref{q/W^2 s=0 gral l})
 to $O(\nb^{15})$ --  see~\cite{Casals:Ottewill:2012SmallMidBC}. 
(c) $s=0$, $\ell=1$ for large  $\nb$; continuous-blue using Eq.(\ref{eq:Leaver-Liu series for Deltagt}); dashed-red using the asymptotics of Eq.(\ref{eq:large_nu}).
}
\label{fig:DeltaG s=l=2 and s=0,l=1}
\end{figure}


\section{Spin-1 QuasiNormal Modes}

We present here an analysis for large-$n$ of the electromagnetic QNMs.
We may find solutions of  Eq.(\ref{eq:radial ODE}) valid for fixed $\rnu \equiv \rb^2\nb/2$ as expansion in powers of
$\nb ^{-1/2}$ as $\psi_i = \suma_{k=0}^\infty \psi_i^{(\iQ)}$, $i=1,2$, starting with the two independent solutions:
$\psi_1^{(0)}(\rnu)=(2/\nu)\sinh \rnu$
and
$\psi_2^{(0)}(\rnu)=\cosh \rnu$.
We may express any  higher order solution in terms of the $0^\mathrm{th}$-order Green function as
\begin{align}
\label{eq:Green0}
\psi_i^{(\iQ)}(\rnu)&=\!\!\int_0^\rnu\! du \frac{\sinh\left(\rnu-u\right)}{(2 u)^{3/2}}
\left\{
\left[8D^2-6D-\lambda\right]\frac{\psi_i^{(\iQ-1)}(u)}{\sqrt\nb}
\right. \nonumber \\ & \left.-
(2u)^{1/2}\left[4D^2 -2D - \lambda\right]\frac{\psi_i^{(\iQ-2)}(u)}{\nb}
\right\}
\end{align}
where $D\equiv u\tfrac{d}{du}$. From this expression, it follows that
 \begin{align} \label{eq:psi t->it}
&
\psi_1^{(\iQ)}(e^{i \pi} \rnu) = -e^{i \pi \iQ/2}  \psi_1^{(\iQ)}(\rnu) 
\\ &
\psi_2^{(\iQ)}(e^{i \pi} \rnu) = e^{i \pi \iQ/2} \left[ \psi_2^{(\iQ)}(\rnu) - \A  \psi_1^{(\iQ-1)}(\rnu)\right]
\nonumber
 \end{align}
 where $\aa\equiv -{\lambda \sqrt\pi}/{2}$. In addition, for $\arg(r)=\pi/4$, 
$e^{-i \pi (\iQ-2)/4}  \psi_1^{(\iQ)}(\rnu) $
and 
$e^{-i \pi \iQ/4} \left[\psi_2^{(\iQ)}(\rnu) +
 \tfrac{1}{2}\A  \psi_1^{(\iQ-2)}(\rnu)\right] $
are both real.
It follows 
that along $\arg (r)=\pi/4$, 
up to power law 
corrections, 
 \begin{align} \label{eq:psi pi/4}
&
\psi_i \sim \Ai \ept  + \Bi \emt 
\\ &
\En\equiv  \frac{1}{\nb} \suma_k \frac{\alpha_\iQ }{\nb^{\iQ/2}},\quad
\Gn \equiv  \frac{1}{2}  \suma_k \frac{\beta_\iQ- \A \alpha_{\iQ-2} }{\nb^{\iQ/2}}
\nonumber\\&
\Fn\equiv -  \frac{1}{\nb}  \suma_{k} \frac{ i^\iQ\alpha_\iQ^*}{\nb^{\iQ/2}} 
, \quad
\Hn \equiv \frac{1}{2} \suma_{k} \frac{ i^\iQ \left(\beta_\iQ^* - \A \alpha_{\iQ-2}^*\right) }{\nb^{\iQ/2} }
\nonumber
 \end{align}
 \fixme{Is there really a need to define $\psio$ and $\psit$?}
 Equating asymptotic expansions at $\arg (r)=3\pi/4$ yields
$\alpha_\iQ\in \mathbb{R}$, $\text{Im} \beta_\iQ = \alpha_1^2 \alpha_{\iQ-2}$ and 
also serves to determine $\text{Re}\beta_{\iQ}$ 
(except when $\iQ=4p-2$ for $p \in \mathbb{N}$, which do not contribute to the QNM condition).

 Along $\arg (\rnu) =\pi/2$ the Green functon in Eq.~(\ref{eq:Green0}) is rapidly oscillating and we can obtain the values of $\alpha_\iQ$  directly 
from a stationary phase analysis~\cite{Bender:Orszag}%
\dropme{ and
 we obtain}:
 \begin{align} \label{eq:alpha values}
 &
 \az=1,\quad 
  \aa=-\frac{\lambda \sqrt\pi}{2},
\quad
  \ab=\frac{\lambda^2 \ln 2}{2}-\frac{\lambda}{12}
 \\ &
 \ac=
\frac{\lambda^3\sqrt{\pi}(      4 \ln 2-\pi) }{8}
-\frac{11\sqrt{\pi}\lambda^2}{48}
+\frac{41
      \sqrt{\pi} \lambda }{192}
+
      \frac{\sqrt{\pi}}{16}   
 \nonumber
\end{align}

\begin{figure}[t]
\includegraphics[width=\columnwidth]{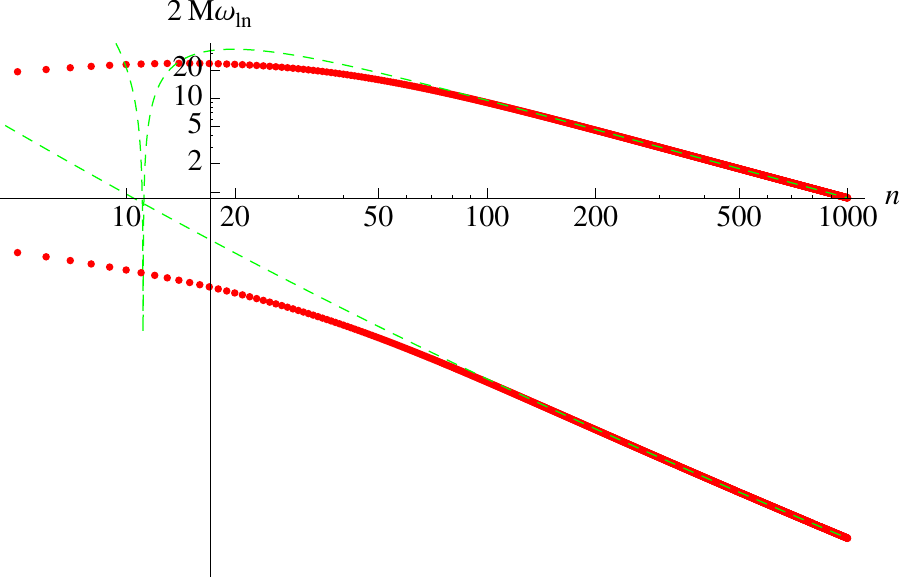}
\caption{
(Color online).
Log-log plot of QNM frequencies for $s=\ell=1$ from the asymptotics Eq.(\ref{eq:QNM s=1}) in dashed-green
and the numerical data in~\cite{QNMBerti} in dotted-red.
The two upper curves correspond to $400\left|\text{Im}(\wbQNM)+\frac{n}{2}\right|$ and the two lower
curves to $\text{Re}(\wbQNM)$.
 }
\label{fig:QNM numeric closed form}
\end{figure}

By matching the $\psi_i$ to WKB solutions along $\arg(r)=\pi/4$ and $3\pi/4$ we are able to find large-$\nu$ asymptotics for $\g$.
Also, we may use the exact monodromy condition, $\f\left((r-r_h)e^{2\pi i},\ob\right)=e^{2\pi \ob}\f(r-r_h,\ob)$, to
obtain large-$\nb$ asymptotics for $\f$.
The asymptotic QNM condition
($W=0$)
in the 4th quadrant then becomes
 \begin{equation}\label{eq:QNM cond}
 e^{-4\pi \nb i}-1=\frac{2\left(\Hn\Fn^*-\Fn\Hn^*\right)}{\left(\En\Hn-\Fn\Gn\right)\Fn} \sum\limits_{k\  \text{odd}} \frac{\alpha_\iQ }{\nb^{\iQ/2}}.
 \end{equation}
\leaveout{
 The QNM frequencies are then given by
  \begin{align} \label{eq:QNM s=1 alpha_i}
& \wbQNM
= -\frac{in}{2}-\frac{2i\alpha_1^2}{\pi n}-
\frac{e^{-i\pi/4}4\alpha_1^3}{\pi n^{3/2}}
+\frac{12\alpha_1^4}{\pi n^2}
\\ &
-\frac{e^{i\pi/4}8 \alpha_1^2 \left(4 \alpha_1^3+\alpha_2 \alpha_1-\alpha_3\right)}{\pi n^{5/2}}
+O\left({n^{-3}}\right)
\nonumber
\end{align}
}
It is straightforward to find the QNM frequencies to  {\it arbitrary} order in $n$ 
in terms of the $\alpha_\iQ$ 
by systematically solving Eq.(\ref{eq:QNM cond}).
Explicitly, using the values in Eq.(\ref{eq:alpha values}), we have
 \begin{align} \label{eq:QNM s=1}
& \wbQNM
= -\frac{in}{2}-\frac{i\lambda^2}{2n}+
\frac{e^{-i\pi/4}\pi^{1/2}\lambda^3}{2n^{3/2}}
+\frac{3\pi\lambda^4}{4n^2}+
\\ &
\frac{e^{i\pi/4} \sqrt{\pi } \lambda ^2   \left[72 \lambda ^3 (\pi +\ln 4)-52 \lambda ^2+41 \lambda +12\right]}{96n^{5/2}}
+O\!\left({n^{-3}}\right)
\nonumber
\end{align}
It is remarkable that the terms in the expansion show the behaviour $e^{i k \pi/4} (\mathbb{R})/n^{k/2}$ to 
all orders. 
In Fig.~\ref{fig:QNM numeric closed form} we compare
these asymptotics
with the numerical data in~\cite{QNMBerti}.
In~\cite{Casals:Ottewill:2011largeBC} we apply the method used to obtain Eq.(\ref{eq:QNM s=1}) to the cases
$s=0$ and $2$
and we obtain the corresponding QNM frequencies up to order $n^{-1/2}$ 
and have agreement with~\cite{Musiri:2003bv,MaassenvandenBrink:2003as}. 
\fixme{Comment as to why closed form seems to do better than asymptotics?}

\fixme{As a matter of principle, one could do this for $s=0,2$?}

\fixme{
It seems that we have a formal expression to all orders for $Aout\sim \cf$ and also for $Ain\propto W$. 
Could we calculate the residues at the QNMs?
If so, either using a similar
one for $\f$ at the QNMs (eg, a closed form for $\f$ is known at the algebraically-special freqs.) or using asymptotics of  $\f$ for $r\to \infty$ 
we could then perhaps calculate the QNM series...?}

\fixme{We can already calculate/plot the BC for $s=1$ to two orders higher}

%
%


\begin{acknowledgments}
We are thankful to Sam Dolan and, particularly, to Barry Wardell for helpful discussions.
A.O. thanks Luis Lehner and the Perimeter Institute for Theoretical Physics for hospitality and financial support.
M. C. is supported by a IRCSET-Marie Curie International Mobility Fellowship in Science, Engineering and Technology.
A.O. acknowledges support from Science Foundation Ireland under grant no 10/RFP/PHY2847.
\end{acknowledgments}


%


\end{document}